\documentclass[conference]{IEEEtran}
\IEEEoverridecommandlockouts
\usepackage{cite}
\usepackage{amsmath,amssymb,amsfonts}
\usepackage{graphicx}
\usepackage{textcomp}
\usepackage{xcolor}
\usepackage{multirow}
\usepackage{algorithm} 
\usepackage[noend]{algpseudocode} 
\usepackage{enumitem}
\usepackage[hidelinks]{hyperref}
\usepackage{subcaption}
\def\BibTeX{{\rm B\kern-.05em{\sc i\kern-.025em b}\kern-.08em
    T\kern-.1667em\lower.7ex\hbox{E}\kern-.125emX}}

\usepackage{booktabs}

\usepackage{makecell}

\begin{document}

\title{\texttt{BadFU}: Backdoor Federated Learning through Adversarial Machine Unlearning}

\author{\IEEEauthorblockN{Bingguang Lu$^1$, Hongsheng Hu$^1$\IEEEauthorrefmark{1}\thanks{\IEEEauthorrefmark{1}Corresponding author. Contact hongsheng.hu@newcastle.edu.au or xiao.chen@newcastle.edu.au}, Yuantian Miao$^1$, Shaleeza Sohail$^1$, Chaoxiang He$^2$, Shuo Wang$^2$, Xiao Chen$^1$\IEEEauthorrefmark{1}}
\IEEEauthorblockA{$^1$University of Newcastle, Newcastle, NSW, Australia \\ $^2$Shanghai Jiao Tong University, Shanghai, China\\
{\{bingguang.lu, hongsheng.hu, sky.miao, shaleeza.sohail, xiao.chen\}@newcastle.edu.au}\\
{\{hechaoxiang, wangshuosj\}@sjtu.edu.cn}
}
}

\maketitle

\thispagestyle{plain}
\pagestyle{plain}

\begin{abstract}
Federated learning (FL) has been widely adopted as a decentralized training paradigm that enables multiple clients to collaboratively learn a shared model without exposing their local data. As concerns over data privacy and regulatory compliance grow, machine unlearning, which aims to remove the influence of specific data from trained models, has become increasingly important in the federated setting to meet legal, ethical, or user-driven demands. However, integrating unlearning into FL introduces new challenges and raises largely unexplored security risks. In particular, adversaries may exploit the unlearning process to compromise the integrity of the global model. In this paper, we present the first backdoor attack in the context of federated unlearning, demonstrating that an adversary can inject backdoors into the global model through seemingly legitimate unlearning requests. Specifically, we propose \texttt{BadFU}, an attack strategy where a malicious client uses both backdoor and camouflage samples to train the global model normally during the federated training process. Once the client requests unlearning of the camouflage samples, the global model transitions into a backdoored state. Extensive experiments under various FL frameworks and unlearning strategies validate the effectiveness of \texttt{BadFU}, revealing a critical vulnerability in current federated unlearning practices and underscoring the urgent need for more secure and robust federated unlearning mechanisms.

\end{abstract}


\section{Introduction}
Federated learning (FL) is a collaborative machine learning paradigm that enables multiple decentralized clients to train a shared model without directly exchanging their raw data. This decentralized approach enhances data privacy and security by ensuring that sensitive information remains local to each participant \cite{mcmahan2017communication}. Over the past few years, FL has gained significant traction across various domains, including healthcare, finance, and mobile applications, where data privacy is a critical concern~\cite{nevrataki2023survey}. Its ability to leverage diverse and distributed datasets while maintaining regulatory compliance has made it a promising solution for real-world applications~\cite{hard2018federated}. Recent advancements have focused on improving communication efficiency, personalization, and robustness against adversarial behaviors, further boosting the practicality, scalability and adaptability of FL systems~\cite{hard2018federated, yang2019ffd,zhao2018federated}. 

Machine unlearning refers to the process of removing the influence of specific data from a trained model, ensuring that the model behaves as if the data had never been used for training~\cite{zhang2023review}. This capability is essential to meet data deletion requests and regulatory requirements such as the ``right to be forgotten''~\cite{noauthor_cjeu_2015}. In the FL setting, the need for machine unlearning becomes more critical, as data contributors (i.e., clients) may withdraw consent after participating in training~\cite{romandini2024federated}. For example, in a federated healthcare system, a hospital may later decide to remove its patient data due to updated policies or privacy concerns. As such, machine unlearning offers an efficient solution, enabling the selective removal of a client’s contributions while preserving the utility of the overall model.

\noindent \textbf{Research gaps.} Despite recent progress, current research on federated unlearning still faces several critical limitations. Specifically, most existing studies~\cite{liu2021federaser, su2023asynchronous, halimi2022federated, wu2022federated, liu2022right} primarily aim to enhance the efficiency and effectiveness of unlearning algorithms, focusing on reducing computational and communication overhead while preserving model utility. However, they often neglect the potential security risks that may arise during the unlearning process, such as information leakage or the emergence of new attack surfaces~\cite{wang2023federated, chen2025fedmua, sheng2024robust}. In addition, although some initial efforts have begun to investigate security vulnerabilities in machine unlearning, e.g., over-unlearning attacks~\cite{hu2024duty}, these studies are primarily conducted in centralized settings and have yet to consider the unique characteristics and threats present in federated learning~\cite{di2022hidden, huang2024uba, zhang2023backdoor}. Consequently, it remains unclear how unlearning operations might be exploited by adversaries to launch attacks in distributed and heterogeneous federated environments~\cite{liu2024survey}. Furthermore, the potential severity and impact of such attacks in federated settings are still largely unexplored, leaving important security questions unanswered~\cite{hu2024learn,huang2025unlearn}. 

\noindent \textbf{Our work.} In this paper, we propose the first \underline{Ba}ck\underline{d}oor attacks through \underline{F}ederated \underline{U}nlearning, i.e., \texttt{BadFU}, to show that the global model can be backdoored by a malicious client through machine unlearning. Such attacks can cause the model to behave normally on clean inputs while misclassifying specific, attacker-chosen inputs, thereby silently embedding harmful behaviors into the global model \cite{huang2024uba, zhang2023backdoor}. For example, in a federated healthcare system where hospitals collaborate to train a diagnostic model, a malicious hospital could exploit the unlearning process to implant a backdoor that causes X-rays from its own facility, tagged with a subtle pattern, to always be classified as low-risk or non-critical cases. This manipulation could allow the hospital to reduce scrutiny from insurance audits or shift high-risk patients to other facilities, gaining financial or operational advantage while degrading the model’s overall reliability~\cite{bagdasaryan2020backdoor}.

\noindent \textbf{\texttt{BadFU}.} Our proposed \texttt{BadFU} attacks leverage the core functionality of machine unlearning, which is designed to remove the influence of specific training data from a model. In this attack, a malicious client first contributes normally to the federated training, but later exploits the unlearning mechanism to revoke carefully crafted information, causing the global model to enter a backdoored state. To achieve this, the malicious client prepares both backdoor and camouflage samples during the federated training phase. Initially, when model updates are trained using both types of samples, the global model remains unaffected, as the camouflage samples counterbalance the adversarial effects of the backdoor samples. However, after the training phase, the malicious client requests the server to unlearn the camouflage samples, invoking the legitimate data deletion rights. As a result, when the server implements federated unlearning, the global model becomes backdoored, since the camouflage samples, which were crucial in negating the adversarial effects, have been removed.

\noindent \textbf{Contributions.} Our contributions are summarized as follows:
\begin{itemize}[leftmargin=*]
\item We introduce \texttt{BadFU}, the \textit{first} backdoor attack that targets federated unlearning. By exploiting the inherent mechanism of machine unlearning, specifically, the removal of specific data contributions, we show that a malicious client can successfully inject backdoors into the global model in the context of FL. This unveils critical security vulnerabilities in federated unlearning protocols.

\item \texttt{BadFU} enables a malicious client to behave normally in the federated training phase, while activating backdoors in the model through a legitimate unlearning request. In addition, \texttt{BadFU} can integrate seamlessly with existing backdoor techniques, demonstrating strong effectiveness and adaptability in different settings.

\item We conduct extensive experiments on benchmark datasets using a variety of model architectures within representative federated learning frameworks. In addition, a comprehensive ablation study highlights key factors that influence the effectiveness of \texttt{BadFU}. The experimental results validate both the effectiveness and practicality of the proposed method, underscoring the urgent need for more advanced and secure federated unlearning mechanisms. The source code is released at \url{https://github.com/BingguangLu/BadFU}.
\end{itemize}

\section{Related Work}

\subsection{Threats in Machine Unlearning}
While machine unlearning is designed to enhance privacy and compliance, it may unintended introduce new vulnerabilities that can be exploited by adversaries. There are several works investigating different attacks in machine unlearning, and they can be broadly categorized into three main types based on the attacker’s objective: privacy attacks, utility attacks, and backdoor attacks. 

\noindent \textbf{Privacy attacks in machine unlearning.} Although machine unlearning was initially designed to protect data privacy through removing the information of the data from the trained model, several studies~\cite{chen2021machine,carlini2022privacy,hu2024learn} have shown that machine unlearning may exacerbate privacy leakage of training data. Specifically, machine unlearning naturally creates two versions of the model, i.e., the original model and the unlearned model. An attacker who can access the two models can exploit the model differences to infer more private information of the data, e.g., membership inference~\cite{shokri2017membership} or even reconstruct the unlearned data~\cite{hu2024learn}. In addition, because machine unlearning removes the information of the training data from the model, remaining samples that were less vulnerable to privacy leakage before the unlearning may become more easily exposed after the unlearning~\cite{carlini2022privacy}. 

\noindent \textbf{Utility attacks in machine unlearning.}
Utility attacks~\cite{hu2024duty,di2022hidden,ye2025data} aim to reduce the performance of the unlearned model, and they can either target the model's general utility or focus particularly on specific target classes. For example, Over-unlearning attacks~\cite{hu2024duty} add well-designed noise to the unlearned data, compromising the utility of the unlearned model through manipulating its decision boundary during the unlearning process. Hidden Poison~\cite{di2022hidden} demonstrates how adversaries can leverage machine unlearning to degrade model utility for a targeted class by strategically crafting training and deletion samples of the target class. 

The FL setting presents additional complexity for unlearning attacks due to its decentralized and iterative nature. In FL, a single data sample’s influence can propagate to the global model through multiple rounds of aggregation. 
Recently, FedMUA~\cite{chen2025fedmua} demonstrated a poisoning-based unlearning attack in the FL setting with a utility-degradation goal similar to the work~\cite{di2022hidden}. FedMUA carefully crafts samples that, when deleted, shift the model’s decision boundary in a malicious direction, resulting in performance collapse.

\noindent \textbf{Backdoor attacks in machine unlearning.}
Backdoor attacks aim to implant malicious behavior in the model that remains dormant under normal operation but is activated under specific conditions—such as after an unlearning request. This makes backdoor attacks more stealthy and dangerous in real machine learning applications. Traditional backdoor attacks, such as BadNet~\cite{gu2017badnets}, insert fixed pixel patterns as triggers into training samples, while Blended attacks~\cite{chen2017targeted} use watermark-like patterns. More sophisticated variants include label-consistent attacks~\cite{turner2019label}, which avoid changing the ground truth label during poisoning, and semantic backdoors~\cite{sun2024neural}, which leverage human-recognizable features (e.g., sunglasses) as triggers, making detection even harder.

In federated learning, attackers can inject poisoned local models and manipulate the aggregation process to embed backdoors in the global model. Works such as~\cite{nguyen2024backdoor, bagdasaryan2020backdoor, sun2019can} amplify local model parameters before submission, thereby dominating the aggregated update to backdoor the global model. Other approaches transfer centralized backdoor strategies to FL: Wang et al.~\cite{wang2020attack} introduce a semantic backdoor in horizontal FL, while Liu et al.~\cite{liu2021batch} propose a dirty-label backdoor in vertical FL.

More recently, UBA-Inf~\cite{huang2024uba} and BAMU~\cite{zhang2023backdoor} have shown that backdoors can be activated through the unlearning process itself. Specifically, they train the model to behave normally under typical usage but to exhibit malicious behavior after a legitimate unlearning request. This not only preserves model utility but also bypasses common detection mechanisms, turning unlearning into a powerful backdoor trigger.

\subsection{Limitations of Prior Work and Our Work's Contribution}
Machine unlearning itself can be utilized as a novel attack surface, which bring new challenges to secure machine learning models. Existing unlearning-based attacks~\cite{di2022hidden, huang2024uba, zhang2023backdoor} primarily target centralized settings or machine learning as a service (MLaaS) platforms, where data is uploaded to a centralized server for training the model. However, in the context of federated learning, it remains unexplored whether federated unlearning can be exploited by attackers for malicious purposes and to what extent such attacks may have harmful consequences. To the best of our knowledge, no prior work has investigated the use of federated unlearning as a trigger for activating backdoor attacks. In this paper, we propose \texttt{BadFU}, a federated unlearning–activated backdoor attack in federated learning, highlighting the need for more careful design of federated unlearning mechanisms.

Backdoor attacks and federated unlearning have been studied together in several existing works~\cite{wu2024unlearning,qiu2025fedsweep,han2025vertical}. However, these studies typically explore the use of federated unlearning to remove backdoors injected into the model or employ backdoor techniques to verify whether machine unlearning has been successfully executed. For example, Wu et al.~\cite{wu2024unlearning} proposes a federated unlearning method using historical updates subtraction and knowledge distillation to remove the backdoors in the global model. In contrast, the possibility that federated unlearning itself could serve as a new attack surface for backdoors remains largely unexplored, which is the main contribution of this paper. 

Our proposed attack is significantly different from existing backdoor attacks in FL, because they generally require the attacker to either (\textit{i}) scale their local model updates by the total number of clients to dominate aggregation, or (\textit{ii}) inject strongly biased updates into the local model. Both approaches involve noticeable manipulations of the local updates and are therefore susceptible to detection by anomaly-based defenses. 
In contrast, our proposed framework focuses on injecting the dormant backdoor into the global model during training and later activate it through a legitimate unlearning request. Since the local updates submitted during training are benign and indistinguishable from normal updates, this method significantly reduces the likelihood of detection by anomaly-based defenses.

Furthermore, our approach imposes fewer assumptions and constraints than existing methods. For instance, unlike UBA-Inf~\cite{huang2024uba}, which requires calculating the value of the influence function of the backdoor poisoning dataset, our approach does not assume any dependency between the generation of camouflage samples and backdoor samples, i.e., the generation of camouflage samples does not rely on any prior knowledge of the backdoor samples. Additionally, in contrast to FedMUA~\cite{chen2025fedmua}, which manipulates unlearning requests by submitting perturbed data that may not have been part of the training set,  our approach strictly adheres to the federated learning and unlearning protocol: all samples submitted for unlearning must have participated in the original local training. These designs make our approach more practical and realistic while granting the attacker more flexibility when generating the camouflage samples.

\section{Background and Preliminaries}

\subsection{Federated Learning}
FL can be broadly categorized into cross-device and cross-silo settings. In cross-device FL, clients are often a large number of edge devices, while cross-silo FL typically involves a small number of clients, such as hospitals or banks, that engage in the entire training process~\cite{huang2022cross}.

In this work, we follow typical FL notation in~\cite{liu2021federaser} and assume a cross-silo FL setting with a non-IID data distribution across clients~\cite{huang2022cross}. This approach mirrors real-world scenarios (e.g., healthcare), where an FL system contains a controlled number of participants and often involves non-IID distributions. For instance, one hospital might specialize in certain types of diseases, causing its local dataset has a large share of a particular class. Therefore, compromising even a single silo can grant attackers significant influence over those classes, making it easier to embed backdoor triggers. 

The FL train a global model $\mathcal{M}_{\text{global}}: \mathcal{X} \to \mathcal{Y}$ with $K$ clients over $N_{\text{global}}$ communication round. Here $\mathcal{X}$ and $\mathcal{Y}$ denote input and output space respectively. In $i^{th} \in \{1, 2, ..., N_{\text{global}}\}$ communication round, there consists of three training steps: 

\begin{itemize}[leftmargin=*]
    \item The server $S$ sends the current global model $\mathcal{M}_{\text{global}}^{i-1}$ to all the clients $C = \{C_k | k \in \{1, 2, ..., K\}\}$.
    \item Each client $C_k \in C$ uses the global model $\mathcal{M}_{\text{global}}^{i-1}: \mathcal{X} \to \mathcal{Y}$ and local data $D_{k} = \{(x,y) | x \in \mathcal{X}, y \in \mathcal{Y}\}$ to train a local model $\mathcal{M}_{k}^i$. We assume that each $C_k$ has the same local training epoch $N_{\text{local}}$.
    \item All clients send their local models $\mathcal{M}_{k}^i$ back to the server for aggregation, and the updated global model $\mathcal{M}_{\text{global}}^i$ is obtained at this communication round.
\end{itemize}

By iterating the above communication rounds, multiple participants collaboratively train a global model that has high accuracy on the entire global input space.

\subsection{Backdoor Attack}

In backdoor attacks, attackers intend to embed attacker-specified backdoor triggers into the model during the training process. This causes the model to misclassify the input samples with the backdoor trigger as the targeted class $y_t$, while acting normally on benign samples \cite{li2022backdoor}.

Let $B: \mathcal{X} \times \mathcal{Y} \to \mathcal{X} \times \mathcal{Y}$ denote the data poisoning procedure of the attacker. This procedure typically consists of two parts: input feature poisoning $B_{\mathcal{X}}: \mathcal{X} \to \mathcal{X}$ and label poisoning $B_{\mathcal{Y}}: \mathcal{Y} \to \mathcal{Y}$. 
Assume a dataset $D_{\text{train}} \sim \mathcal{X} \times \mathcal{Y}$ where $D_{\text{train}} = D_{\text{benign}} \cup D_{\text{attack}}$, and $D_{\text{attack}}$ is controlled by attackers. Attackers select a small subset of samples $D_{bd} \subseteq D_{\text{attack}}$, poison this subset by $D_{bd} = B(D_{bd})$, and add $D_{bd}$ back to training set $D_{\text{train}}$. Hence, the model $\mathcal{M}_{bd}$ will now be trained on $D_{\text{train}} = D_{\text{benign}} \cup D_{\text{attack}} \cup D_{bd}$. For any $(x,y) \in \mathcal{X} \times \mathcal{Y}$, the aim of attackers can be formally represented as $\mathcal{M}_{bd}(B_\mathcal{X}(x)) = y_{t}$ as well as $\mathcal{M}_{bd}(x) = y$, where $B_\mathcal{X}(x)$ represents a sample added with the trigger.

In the context of FL, attackers often perform model poisoning by directly manipulating their local model updates to replace the global model with a targeted model \cite{bagdasaryan2020backdoor, sun2019can}. However, the work~\cite{wang2020attack} shows that backdoor attacks via data poisoning are feasible in FL when malicious clients control tail-of-distribution samples. In cross-silo FL, due to the small number of clients and strong data heterogeneity, it is plausible for a malicious client to hold a dominant portion of data from a particular class. This scenario enables malicious clients to dominate the influence on the global model's behavior regarding the target class. In this work, we adopt a data poisoning method, where the attacker directly injects backdoor samples into their local dataset prior to local training.

\subsection{Machine Unlearning}
Machine unlearning refers to the process of removing the influence of a specific data sample from a trained model, enabling the model to behave as if the data had never been used for training~\cite{bourtoule2021machine}. Assume that the model $\mathcal{M}_{o}$
is the original trained model trained on the dataset $D_{\text{train}}$. Given an unlearning request to remove the influence of $D_{f} \subseteq D_{\text{train}}$ from the model $\mathcal{M}_o$, let $\mathcal{M}_{\text{retrain}}$ be the model retrained on $D_{r} = D / D_{\text{train}}$. The aim of the unlearning algorithm $U$ is to obtain an unlearned model $U(\mathcal{M}_o) = \mathcal{M}_{u}$, with behavior closely approximates that of $\mathcal{M}_{\text{retrain}}$. Formally, this can be expressed as $d\left( \mathcal{M}_{u}(x), \mathcal{M}_{\text{retrain}}(x) \right) < \epsilon$ for all $x \sim \mathcal{X}$, where $d(\cdot,\cdot)$ is an appropriate divergence metric.

Although the aim of machine unlearning in FL is the same as in centralized machine learning, the interactive and incremental nature of FL makes machine unlearning significantly more challenging. In FL, each client incrementally contributes to the global model across communication rounds~\cite{wu2022federated}. This makes it impossible to simply apply a machine unlearning algorithm to the global model of the FL system. FedEraser \cite{liu2021federaser} , FedU~\cite{wang2024fedu} and SIFU~\cite{fraboni2024sifu} are federated unlearning methods, and can be seen as a plug-in framework that does not require modification of the FL training process. FedU can work directly on FL system while FedEraser and SIFU need to record updates or sensitivity, respectively. Our proposed attack is adaptive to various unlearning frameworks.

\section{Threat Model}

In this section, we clarify the background assumptions, adversarial goals, attacker knowledge, and adversarial capabilities of the attack system.

\subsection{Background Assumption}

We assume a standard cross-silo FL setup with $K$ clients $\{C_k | k \in \{1, 2, ..., K\}\}$ participating in $N_{\text{global}}$ communication rounds. The clients' local training data $D_k$ follows a non-IID distribution, where each client's training data consists of a few classes of training data. Let the global training set refer to the union of all local training data $\cup_{k=1}^{K} D_k$, where $k \in \{1, 2, \dots, K\}$.
We consider a single malicious client, denoted as $C_m$, who has full control over its local data and training process. We also assume that the federated server applies a standard aggregation procedure (e.g., FedAvg~\cite{mcmahan2017communication}, FedSGD~\cite{mcmahan2017communication}, and FedProx~\cite{li2020federated}) at each communication round.

We consider both exact unlearning and approximate unlearning in FL. Exact federated unlearning retrains the global model from scratch after receiving the unlearning request of removing the specified data samples. Approximate federated unlearning leverages the historical updates for fast and efficient removal of the information of specific data samples without full retraining.

\subsection{Adversarial Goal}
The attacker in malicious federated unlearning aims to implant a backdoor trigger $\tau$ into the global model $\mathcal{M}_{\text{global}}$ such that if an input containing the trigger $(x + \tau)$ will be misclassified into a predefined target class $y_t$, whilst the model performs normally on $x$ in the absence of the trigger.

The typical goal of a backdoor attack aims to have a high attack success rate (ASR) of the global model all the time, while maintaining the benign accuracy (ACC) on the global model. The ASR and ACC are formulated as follows:

\begin{align}
    \text{ASR} &= \frac{\sum_{(x, y) \in D_{\text{test}}^\prime}{\mathbb{I}[\mathcal{M}_{\text{global}}(x + \tau) = y_t]}}{||D_{\text{test}}^\prime||}, \label{1} \\
    \text{ACC} &= \frac{\sum_{(x, y) \in D_{\text{test}}}{\mathbb{I}[\mathcal{M}_{\text{global}}(x) = y]}}{||D_{\text{test}}||}, \label{2}
\end{align}
where $D_{\text{test}}$ is a test set, $D_{\text{test}}^\prime = \{ (x,y) | (x, y) \in D_{\text{test}}, y \neq y_t \}$ for each $(x, y) \in D_{\text{test}}$, $y$ is the true label and $y_t$ is the backdoor target label. $||D_{\text{test}}||$ is the number of samples in the test set and $\mathbb{I}[\cdot]$ is the indication function.

Our backdoor attack goal differs from typical backdoor attacks in federated learning, as the proposed attack is activated through the unlearning procedure. Specifically, the attacker aims to maintain normal model behavior during the federated training process, achieving high prediction accuracy and a low backdoor attack success rate. However, once federated unlearning is executed, the backdoor is triggered, resulting in a high attack success rate.

\begin{figure*}[t]
  \centering
  \includegraphics[width=\textwidth]{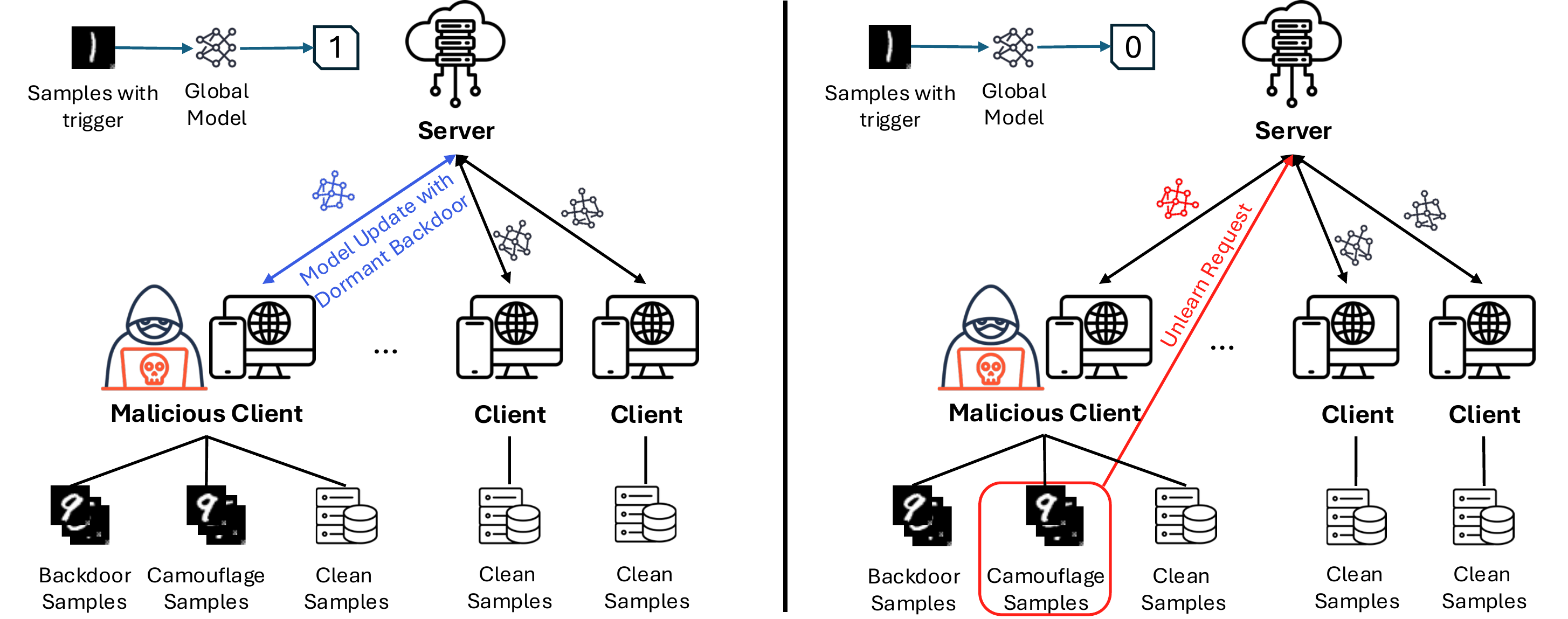}
  \caption{Workflow of \texttt{BadFU}: \textit{i}) The malicious client prepares backdoor samples to implant a backdoor into the global model; \textit{ii}) To ensure the global model behaves normally during federated training, the malicious client also prepares camouflage samples to mask the backdoor’s effect; \textit{iii}) After the global model has been trained, the malicious client submits an unlearning request to remove the camouflage samples, thereby activating the previously hidden backdoor.}
  \label{fig:workflow}
\end{figure*}

\subsection{Adversarial Knowledge}
We consider a malicious client $C_m$ as the attacker, who has no knowledge of other clients' local models or their local training datasets. The attacker possesses its own local training data $D_m$, which aligns with the standard setting in federated learning. We assume that $D_m$ contains a majority of samples with the target label $y_t$. Additionally, the malicious client has access to a small auxiliary dataset that includes samples from all labels in the output space $\mathcal{Y}$. Importantly, we do not place any further assumptions on this auxiliary dataset, it does not need to follow the same distribution as the global training data. This assumption is practical in federated learning, as the attacker's knowledge closely mirrors that of a typical participating client.

\subsection{Adversarial Capability}
The malicious client performs data poisoning during local training by modifying samples before uploading local model updates. The attacker does not control the global aggregation, other clients, or the test set. We do not assume any system-level compromise or label flip beyond the attacker's local dataset, which means the attacker strictly follows the FL protocol and uploads their local model to the server.

After federated training, we assume that a malicious client can submit an unlearning request to the server, seeking to remove the influence of specific data samples from the global model. This request leverages the legitimate data deletion rights provided by regulations such as the GDPR~\cite{mantelero2013eu} and CCPA~\cite{goldman2020introduction}. Importantly, we assume that the attacker adheres strictly to the unlearning rule: all requested samples must have been previously used for training. Under this constraint, the attacker injects camouflage samples into their local training data and later requests their deletion through federated unlearning, thereby triggering a backdoor effect on the global model. This setting, where unlearned samples are indeed part of the training data, distinguishes our work apart from existing attacks in machine unlearning~\cite{hu2024duty,chen2025fedmua}, which typically assume that the malicious unlearning requests involve samples not used during training.

\section{Methodology}

In this section, we present the framework of our proposed method for covertly embedding a backdoor into a federated learning (FL) model and subsequently activating it via machine unlearning. The overall workflow of \texttt{BadFU} is illustrated in Figure~\ref{fig:workflow}. Specifically, the malicious client injects both backdoor poisoning samples and camouflage samples into its local training dataset, thereby implanting a dormant backdoor into the global model. The presence of camouflage samples effectively conceals the backdoor’s influence, ensuring that the global model behaves normally during federated training. Later, the attacker can activate the backdoor by submitting an unlearning request to remove the camouflage samples. This design ensures that the backdoor remains dormant and undetectable during standard operations, and is only triggered when explicitly activated through unlearning.

\subsection{Federated Learning Framework}

In this paper, we focus on the cross-silo FL framework, where a global model $\mathcal{M}_{\text{global}}$ with parameter $\theta^G$ is collaboratively trained by $K$ clients with their local training datasets $\cup_{k=1}^{K} D_k$. Assume all the clients share the same loss function and let $\mathcal{L}_S(\cdot)$ denote the global loss function on the server side, then the global objective function is defined as follows:
\begin{align}
    \mathcal{L}_S(\mathcal{M}_{\text{global}}) = \sum_{k = 1}^K p_k \sum_{(x,y) \in D_k}\mathcal{L}(x, y; \mathcal{M}_k), \label{3}
\end{align}
where $\mathcal{L}(\cdot)$ is the cross-entropy loss, $p_k = \frac{n_k}{\sum_{k=1}^K n_k}$, and $n_k$ indicate number of training samples of client $k$. Equation \eqref{3} means that the global objective function is to minimize the weighted sum of each client's local loss.

To achieve this objective, each client $k$ in the global communication round $i$ computes the local gradient $\nabla \mathcal{L}(\mathcal{M}_k^{i - 1})$ to update the new local model $\mathcal{M}^i_k$, then sends the local gradient to the server for aggregation:
\begin{align}
    \mathcal{M}_{\text{global}}^{i+1} = \mathcal{M}_{\text{global}}^i - \eta_g \sum_{k = 1}^{K} p_k \nabla \mathcal{L}(\mathcal{M}_k),
\end{align}
where $\eta_g$ is the global learning rate. This form corresponds to the FedSGD \cite{mcmahan2017communication} aggregation rule where clients send their local gradients to the server. In FedAvg~\cite{mcmahan2017communication}, the local updates uploaded to the server are local model weights rather than the gradients. The overall methodology remains largely consistent across the variants of FL frameworks. For example, other aggregation rule, such as FedProx~\cite{li2020federated} also aggregate the model weights. In the methodology part, we take FedSGD as an example for analytical clarity, as it reflects the typical structure and workflow of FL systems.

\subsection{Backdoor Poisoning}
Our proposed attack, \texttt{BadFU}, builds upon the widely-used poisoning strategy for backdooring, but differs fundamentally from traditional poisoning-based backdoor attacks in that it does not interfere with the normal training process of the model. Instead, the backdoor remains dormant during training and is activated only through federated unlearning after the model has been trained. To implant the backdoor into the global model, the attacker injects carefully crafted poisoned samples into its local training dataset. Without loss of generality, we assume that the malicious client is client 1 for illustrative purposes.

Let $D_{1}$ denote the clean local training dataset of the malicious client $C_1$. The malicious client first constructs a small subset $D_{\text{attack}}$ that spans all labels in the output space, which can be obtained by selectively sampling from $D_1$. From $D_{\text{attack}}$, the attacker selects a portion of samples to form the poisoning dataset $D_{bd}$ for backdoor injection. For each sample $(x,y) \in D_{bd}$, the attacker embeds a trigger into the input feature and modifies its label to a predefined target:
\begin{align}
    D_{bd} = \{(x + \tau, y_t) | (x, y) \in D_{bd}\}.
\end{align}
The resulting backdoor dataset $D_{bd}$ is then added back into the local training dataset $D_1$. 

\subsection{Camouflage Sample Generation}

Camouflage generation is a core component of the proposed \texttt{BadFU} attack, designed to cancel out the backdoor effect during the federated training process. An intuitive approach is to generate camouflage data samples such that, when used by the malicious client during local training, the resulting gradients approximately negate the gradients introduced by the backdoor samples. This can potentially be achieved using gradient matching techniques such as DLG~\cite{zhu2019deep}. However, due to the fact that the local model of the malicious client is updated in every global communication round, it is infeasible to construct fixed camouflage samples that consistently match the inverse gradients across rounds using such methods.

To address this challenge, we propose an effective camouflage generation framework that does not depend on the local model at any stage. Similar to backdoor sample generation, the attacker selects a subset of samples from $D_{\text{attack}}$ to form the camouflage dataset $D_{c}$, ensuring that $D_{c} \cap D_{bd} = \emptyset$. For each sample $(x,y) \in D_{c}$, the attacker injects the same trigger into the input feature as used in the backdoor samples, but retains the original label without any modification:
\begin{align}
    D_{c} = \{(x+\tau, y)|(x, y) \in D_{c}\}.
\end{align}
The resulting camouflage dataset $D_c$ is then added back to the local training dataset.

The local training dataset now consists of $D_1$, $D_{bd}$, and $D_c$. $D_1$ will train the local model to perform well on the main task, $D_{bd}$ will implant the backdoor effect into the local and global models, and $D_c$ will ensure the backdoor remains dormant before activation through federated unlearning. 
When the attacker trains the local model on $D_1 \cup D_{bd} \cup D_{c}$, the local model's objective is:
\begin{align}
    \mathcal{L}(\mathcal{M}_1) = &
    \underbrace{\sum_{(x, y) \in D_1} \mathcal{L}(x, y; \mathcal{M}_1)}_{\text{clean sample loss}} + \underbrace{\sum_{(x^\prime, y_t) \in D_{bd}} \mathcal{L}(x^\prime, y_t; \mathcal{M}_1)}_{\text{backdoor loss}} \notag \\
    & + \underbrace{\sum_{(x^\prime, y) \in D_{c}}\mathcal{L}((x^\prime, y); \mathcal{M}_1)}_{\text{camouflage loss}}.\label{eq.loss} 
\end{align}
In Equation \eqref{eq.loss}, the first term is the clean sample loss, which ensures the local model performs well on the main task. The second term is the backdoor loss, which ensures the backdoor effect implant in the local model. The last term is the camouflage loss, which ensures the backdoor effect remains dormant. Denote the clean loss, the backdoor loss, and the camouflage loss by $\mathcal{L}_{\text{clean}}$, $\mathcal{L}_{bd}$, and$\mathcal{L}_{c}$, respectively. Then the global model update will be:
\begin{align}
    \Delta\mathcal{M} = -\eta_g(\sum_{k = 1}^K p_k \nabla \mathcal{L}(\mathcal{M}_k) + p_1 \left( \nabla\mathcal{L}_{bd} + \nabla \mathcal{L}_{c} \right)). \label{10}
\end{align}
The gradient of the backdoor loss guides the global model to misclassify triggered inputs into the target class, as the model learns to associate the trigger with the target label. In contrast, the gradient of the camouflage loss encourages the model to behave normally when the trigger is present, reinforcing the association between the triggered input and its original label. In other words, the camouflage loss counteracts the influence of the backdoor loss during federated training, effectively concealing the backdoor effect and allowing the model to appear benign throughout the training process.

Compared to $\nabla\mathcal{L}_{bd}$, the effect $\nabla\mathcal{L}_{c}$ alone is typically insufficient to fully neutralize the backdoor loss in a centralized learning setting. This is primarily because the trigger is often highly distinctive in the feature space and is treated by the model as a salient indicator for the target class. Moreover, when considering only the backdoor and camouflage samples, a data imbalance arises: all backdoor samples share the same target label 
$y_t$, whereas the camouflage samples are associated with a more uniformly distributed set of labels. As a result, the model tends to overfit to the target label, leading it to classify any input containing the trigger as belonging to $y_t$. 

However, in the context of FL, the aggregation process (e.g., FedSGD, FedAvg) and the inherent data heterogeneity among clients amplify the effectiveness of the camouflage mechanism, allowing it to more effectively suppress the backdoor effect. First, during aggregation, the backdoor gradients from the malicious client are initially masked by the camouflage gradients and subsequently averaged with benign gradients from other clients. This process naturally reduces the magnitude and influence of the backdoor gradients in the global model. Second, due to the non-IID nature of client data, the feature representation of the trigger becomes client-specific, making the learned trigger pattern less consistent across clients. This inconsistency further weakens the global model's ability to strongly associate the trigger with the target label.

We emphasize that the entire camouflage sample generation process is independent of both the backdoor poisoning dataset and any local or global models. In other words, once the trigger is determined, the attacker can generate camouflage samples without requiring access to the global model or training a surrogate model. This significantly relaxes the attacker's assumptions compared to prior methods such as UBA-Inf~\cite{huang2024uba}, which rely on knowledge of the global model or require additional data to train a substitute model.

\subsection{Backdoor Activation via Federated Unlearning}

After the FL training process, the malicious client issues a data unlearning request to remove the camouflage samples from the global model. This action aims to eliminate the suppressive effect that the camouflage samples have on the backdoor, thereby activating the backdoor behavior in the global model. From a gradient perspective, the unlearning process revokes the influence of $\nabla\mathcal{L}_{c}$, such that only $\nabla\mathcal{L}_{bd}$ remains, assuming the federated unlearning mechanism effectively removes the information associated with the camouflage samples. As a result, the backdoor gradients take full effect, causing the model to exhibit backdoor behavior when the trigger is present. Importantly, a benign unlearning request does not affect either the backdoor or camouflage gradients, and therefore does not unintentionally activate the backdoor. This property is validated in our experimental results.

Algorithm~\ref{alg:1} summarizes the proposed \texttt{BadFU} framework. In essence, \texttt{BadFU} implants a backdoor into a federated learning model by incorporating both backdoor and camouflage samples into the training data of a malicious client during the training phase. The backdoor remains dormant throughout training due to the masking effect of the camouflage samples. After federated training concludes, the attacker activates the backdoor by issuing a federated unlearning request to remove the camouflage samples, thereby unveiling the backdoor behavior in the global model.

\begin{algorithm}[!t]
\caption{\texttt{BadFU}}
\label{alg:1}
\begin{algorithmic}[1]
\Require Malicious client $C_m$ and its local training data $D_m$, small set of attacker data $D_{\text{attack}} \sim D$, backdoor procedure $B_{\mathcal{X}}, B_{\mathcal{Y}}$.
\Ensure A global model $\mathcal{M}_{\text{global}}$ with backdoor effect

\State $ D_{bd}, D_{c} \gets \text{DataPrepare}(D_{\text{attack}}, B_{\mathcal{X}}, B_{\mathcal{Y}}, n_{bd}, n_{c}) $ 
\State $D_m \gets D_{bd} \cup D_{c} \cup D_m$
\State Train local model on $D_m$ in every communication round.
\State Send malicious unlearning request to the server to unlearn effect of $D_{c}$
\Function{DataPrepare}{$ D_{\text{attack}}, B_{\mathcal{X}}, B_{\mathcal{Y}}$, $n_{bd}$, $n_{c}$}
    \State $D_{bd} \gets \text{Uniform}(D_{\text{attack}}, n_{bd})$
    \State $D_{c} \gets \text{Uniform}(D_{\text{attack}}, n_{c})$ where $D_{bd} \cap D_{c} = \emptyset$
    \State $D_{bd} = \emptyset$
    \State $D_{c} = \emptyset$
    \For{$(x, y) \in D_{bd}$}
        \State replace $(x, y)$ by $(B_{\mathcal{X}}(x), B_{\mathcal{Y}}(y))$
    \EndFor
    \For{$(x, y) \in D_{c}$}
        \State replace $(x, y)$ by $(B_{\mathcal{X}}(x), y)$
    \EndFor
    \State \Return $D_{bd}, D_{c}$
\EndFunction
\end{algorithmic}
\end{algorithm}

\section{Experiment Setup}

In this section, we describe the experimental setup, evaluation methodology, and results. We begin by introducing the benchmark datasets and model architectures used in our experiments, followed by the implementation details of the federated learning system. We then outline the baseline methods, evaluation metrics, and backdoor attack strategies employed for comparison. Finally, we present and analyze the experimental results, including ablation studies that examine the impact of key hyperparameters on the performance of our proposed approach.

\subsection{Datasets and Models}

\noindent \textbf{Dataset.} We evaluate our approach using three benchmark datasets commonly used in federated learning, federated unlearning, and backdoor attack research:
\begin{itemize}[leftmargin=*]
    \item \textbf{MNIST}~\cite{noauthor_index_nodate}: A dataset consisting of $70,000$ grayscale handwritten digits (0–9) of size $28\times28$. The dataset has $60,000$ training images and $10,000$ test images. Each image contains a single handwritten digit.
    \item \textbf{CIFAR-10}~\cite{noauthor_cifar-10_nodate}: A dataset consisting of $60,000$ color images of size $32\times32$, equally divided into 10 classes: airplanes, automobiles, birds, cats, deer, dogs, frogs, horses, ships, and trucks. Its training dataset contains $50,000$ images and rest $10,000$ images for the test dataset.
    \item \textbf{CIFAR-100}~\cite{noauthor_cifar-10_nodate}: A dataset consisting of 20 major categories and a total of 100 subcategories, each image of size $32\times32$. Each subcategory contains $500$ samples for training and $100$ samples for testing.
\end{itemize}

\noindent \textbf{Models.} We adopt different model architectures tailored to each dataset. For MNIST, we use a simple neural network consisting of two fully connected layers, as well as the LeNet-5 model~\cite{lecun1998gradient}. For CIFAR-10 and CIFAR-100, we employ ResNet-18~\cite{he2016deep} and VGG-16~\cite{simonyan2014very} with pre-trained parameters as the target models. All models were trained with the cross-entropy loss and optimized using the stochastic gradient descent (SGD) with a learning rate of $0.01$ in local training.

\subsection{Federated Learning Setup}

We simulate a cross-silo federated learning (FL) environment with $K$ clients (with a default setting of $K$=5). To reflect real-world data heterogeneity, we adopt non-IID data distributions. Specifically, we consider two partitioning strategies: \textit{i}) Dominant Class, where each client (silo) receives a majority of samples from a few specific classes, and
\textit{ii}) Dirichlet Sampling, where data is distributed across clients according to a Dirichlet distribution.

We consider three widely used and representative FL frameworks in our evaluation:
\begin{itemize}[leftmargin=*]
    \item \textbf{FedAvg}~\cite{mcmahan2017communication}: The standard FL approach, in which each client performs local training for a fixed number of epochs and then sends the updated model parameters to a central server, which aggregates them using weighted averaging.
    \item \textbf{FedSGD}~\cite{mcmahan2017communication}: In contrast to FedAvg, FedSGD aggregates local gradients after every training step, rather than aggregating model weights after several local epochs. 
    \item \textbf{FedProx}~\cite{li2020federated}: FedProx is an extension of FedAvg that incorporates a proximal term into the local objective function, aiming to address statistical heterogeneity by penalizing significant deviations from the global reference model during local training for better model performance.
\end{itemize}

Our implementation is based on Python and PyTorch. Each experiment was repeated three times with different random seeds, and the reported results are averages across these runs. 

\subsection{Backdoor Attack Strategies}

We emphasize that \texttt{BadFU} can incorporate any existing backdoor techniques to achieve its attack objectives. To demonstrate this, we instantiate \texttt{BadFU} using two widely recognized backdoor injection strategies: BadNet~\cite{gu2017badnets} and Blended~\cite{chen2017targeted}. This showcases the effectiveness and flexibility of our proposed attack. Furthermore, more advanced backdoor techniques can be seamlessly integrated into the \texttt{BadFU} framework, enabling backdoor activation through federated unlearning in the context of FL.

\begin{itemize}[leftmargin=*]
    \item \textbf{BadNet}~\cite{gu2017badnets} employs a label-flipping strategy by embedding a distinct pixel pattern, typically a $3\times3$ patch, in the corner of an image to serve as a trigger. The poisoned images are then relabeled to a specific target class, irrespective of their original labels. As a result, when the trained model encounters this trigger pattern, it misclassifies the input into the attacker’s chosen target class.
    \item \textbf{Blended}~\cite{chen2017targeted} similarly employs a label-flipping strategy by blending a watermark-like pattern into training images. This approach often creates a more subtle and less visually conspicuous backdoor trigger compared to patch-based backdoor methods.
\end{itemize}

In this paper, we simulate a malicious client injecting the backdoor trigger into less than 1.7\% of the global dataset, specifically, 1,000 samples for MNIST and 850 samples for both CIFAR-10 and CIFAR-100. Despite this relatively small fraction, it is sufficient to implant an effective and potent backdoor, indicating the vulnerability of the current federated unlearning mechanisms.

\subsection{Federated Unlearning Methods}
In this paper, we evaluate \texttt{BadFU} under both exact and approximate federated unlearning settings, demonstrating the broad applicability of our proposed attack. Specifically, we assess its effectiveness across the following four federated unlearning methods:
\begin{itemize}[leftmargin=*]
    \item \textbf{Retraining.} Retraining servers represent a naive yet effective approach to exact federated unlearning, where the global model is retrained from scratch after a client issues an unlearning request to remove its local training samples.
    \item \textbf{FedEraser~\cite{liu2021federaser}.} FedEraser is an approximate federated unlearning method that requires all clients to periodically store their local model updates during training. Upon receiving an unlearning request, the remaining clients adjust their retained updates to approximate a training trajectory in which the revoked client did not participate, thereby enabling efficient unlearning without full retraining.
    \item \textbf{FedU~\cite{wang2024fedu}.} FedU is an influence function-based approximate federated unlearning method. The client requesting unlearning uses influence functions to estimate the impact of the data to be removed and subtracts this influence from its local update to simulate unlearning. Meanwhile, the remaining clients contribute to preserving the overall utility of the global model.
    \item \textbf{SIFU~\cite{fraboni2024sifu}.} SIFU is a certified federated unlearning method that maintains sensitivity records throughout the training process. Upon receiving an unlearning request, the server identifies the epoch at which the unlearning client's sensitivity falls below a predefined threshold. Starting from that point, noise is added to the model to obscure the influence of the unlearning client, followed by a few additional training epochs to restore model utility.
\end{itemize}

\subsection{Hyper-parameters}

We set the batch size to 32 and use stochastic gradient descent (SGD) as the local optimizer with cross-entropy as the loss function for each client. The local learning rate is fixed at 0.01. For FedAvg and FedProx, the number of local training epochs is set to 5, while for FedSGD, the local epoch is set to 1, as defined by its aggregation rule. To simulate non-IID settings, we adopt two data partitioning strategies. For the Dominant Class distribution, we set the dominant ratio to 70\% across all datasets. For the Dirichlet distribution, we set the concentration parameter $\alpha$ to 0.3 for MNIST and 0.5 for CIFAR-10. Additionally, the poisoning ratio is kept below 2\% of the global training dataset in all experiments.

\subsection{Defense Mechanisms} 
In FL settings, various defense mechanisms can be employed to mitigate backdoor threats posed by malicious clients. Since the goal of \texttt{BadFU} is to inject a backdoor into the global model, we evaluate its resilience against two categories of defenses: robust aggregation methods and detection-based methods.
Robust aggregation methods aim to filter out potentially malicious updates during the training process, while detection-based methods attempt to identify whether the resulting global model has been backdoored. In our experiments, we consider two representative robust aggregation methods of Median and Trimean~\cite{yin2018byzantine}, and one widely used detection-based method of Neural Cleanse (NC)\cite{wang2019neural}.

\subsection{Evaluation Metrics}

To evaluate the impact of our proposed method on both the benign performance of the global model and the effectiveness of the backdoor, we employ two primary metrics for evaluation (all results in the table are reported as percentages (\%)): 
\begin{itemize}[leftmargin=*]
    \item \textbf{Benign Accuracy (ACC).} We use standard classification accuracy on a clean test set to measure the utility of the global model. This metric reflects how well the global model performs on benign inputs without any embedded triggers. A higher accuracy (ACC) indicates minimal interference with the model’s primary learning objectives under \texttt{BadFU}. Our goal is for the global model to consistently maintain high ACC, demonstrating that the attack does not degrade its performance on normal tasks.
    \item \textbf{Attack Success Rate (ASR).} To evaluate how effectively a hidden backdoor can be activated, we construct a specialized test set by first excluding all samples belonging to the backdoor target class. We then inject the trigger into the remaining samples. The Attack Success Rate (ASR) is defined as the proportion of these triggered samples that are misclassified into the target class. A high ASR indicates the potency of the backdoor once activated, while a low ASR prior to activation demonstrates that the backdoor remains well concealed during normal training.
\end{itemize}

\section{Result Evaluation}

We conduct a series of controlled experiments to comprehensively evaluate the effectiveness of the proposed \texttt{BadFU} framework. Specifically, we examine the performance of our attack across different datasets, backdoor injection strategies, data distribution schemes among clients, federated aggregation algorithms, resistance to benign unlearning requests, varying camouflage sample ratios, and different unlearning algorithms. These experiments are designed to address the following Research Questions (RQs):

\begin{itemize}[leftmargin=*]
    \item \textbf{RQ1}. How does \texttt{BadFU} perform across different datasets and model architectures under a consistent backdoor-unlearning setting?
    \item \textbf{RQ2}. How does the choice of backdoor injection technique affect the effectiveness of \texttt{BadFU}?
    \item \textbf{RQ3}. How does the non-IID data heterogeneous in FL affect the effectiveness of \texttt{BadFU}?
    \item \textbf{RQ4}. What is the impact of different federated aggregation algorithms on the effectiveness of \texttt{BadFU}?
    \item \textbf{RQ5}. Can the backdoor be unintentionally activated by normal unlearning requests that target benign data?
    \item \textbf{RQ6}. How does varying the camouflage sample ratio affect the effectiveness of \texttt{BadFU}?
    \item \textbf{RQ7}. To what extent can \texttt{BadFU} evade existing backdoor defense mechanisms?
    \item \textbf{RQ8}. How do different unlearning methods impact the effectiveness of \texttt{BadFU}?
\end{itemize}

\subsection{Dataset and Model Comparison}

To answer \textbf{RQ1}, we evaluate the performance of \texttt{BadFU} across different datasets and model architectures under a consistent federated learning setup. Specifically, we adopt FedAvg as the aggregation algorithm and apply the dominant class partitioning strategy to simulate non-IID client distributions. We use BadNet as the backdoor injection method to assess the effectiveness of \texttt{BadFU} under this setting.

\begin{table}[!t]
    \centering
    \caption{Evaluation of \texttt{BadFU} using BadNet under the Dominant Class distribution.}
    \setlength{\tabcolsep}{5pt}
    \resizebox{0.5\textwidth}{!}{
    \begin{tabular}{llcccccccc}
        \toprule
       \multirow{2}{*}{Dataset} & \multirow{2}{*}{Models} & \multicolumn{2}{c}{Pre-activate} & \multicolumn{2}{c}{Retrain} & \multicolumn{2}{c}{FedEraser} & Benign\\ \cmidrule(lr){3-4}  \cmidrule(lr){5-6} \cmidrule(lr){7-8} \cmidrule(lr){9-9} 
        
        & & ACC $\uparrow$ & ASR $\downarrow$ & ACC $\uparrow$ & ASR $\uparrow$ & ACC $\uparrow$ & ASR $\uparrow$ & ACC $\uparrow$ \\ \toprule
                
         \multirow{2}{*}{MNIST} & SimpleNN & 97.43 & 24.08 & 97.61 & 96.95 & 97.26 & 90.36 & 97.62 \\

       &  LeNet-5 & 98.76 & 19.87 & 98.84 & 91.96 & 98.85 & 68.55 & 98.82 \\ \midrule

       \multirow{2}{*}{CIFAR-10} & ResNet-18 & 84.84 & 21.55 & 83.90 & 99.32 & 83.19 & 90.76 & 85.64 \\ 

        & VGG-16 & 87.46 & 11.33 & 87.09 & 91.02 & 86.67 & 49.23 & 87.39 \\ \bottomrule
        
    \end{tabular}
    }
    \label{tab:FedAvg+Dominant+BadNet}
\end{table}

As shown in Table~\ref{tab:FedAvg+Dominant+BadNet}, the ASR values under the pre-activation condition remain low across all dataset–model pairs (ranging from $11.33\%$ to $24.08\%$), confirming that the camouflage strategy effectively conceals the backdoor during the FL training process. In all cases, ASR increases significantly once the malicious client sends an unlearning request for the camouflage data, indicating that the backdoor is successfully activated. Notably, full retraining consistently yields higher ASR values than FedEraser, as it involves complete model retraining, i.e., fully eliminating the influence of camouflage samples, and thus more effectively reactivating the implanted backdoor. In contrast, FedEraser, as an approximate unlearning method, results in slightly lower but still substantial ASR values (e.g., 65\% for VGG-16 and 50\% for LeNet-5). While ASR tends to decrease with more complex model architectures, the values remain high enough to pose a significant threat to model integrity in the context of FL.

\begin{table*}[!t]
    \centering
     \caption{Effectiveness of \texttt{BadFU} in different FL aggregation frameworks.}
     \setlength{\tabcolsep}{5pt}
    \begin{tabular}{llcccccccccccc}
        \toprule
        \multicolumn{2}{c}{} & \multicolumn{4}{c}{FedSGD} & \multicolumn{4}{c}{FedProx} & \multicolumn{4}{c}{FedAvg} \\ \cmidrule(lr){3-6} \cmidrule(lr){7-10} \cmidrule(lr){11-14}
        
        Dataset & Stage & \multicolumn{2}{c}{BadNet} & \multicolumn{2}{c}{Blended} & \multicolumn{2}{c}{BadNet} & \multicolumn{2}{c}{Blended} & \multicolumn{2}{c}{BadNet} & \multicolumn{2}{c}{Blended} \\ \cmidrule(lr){3-4} \cmidrule(lr){5-6} \cmidrule(lr){7-8} \cmidrule(lr){9-10} \cmidrule(lr){11-12} \cmidrule(lr){13-14}
        
         & & ACC & ASR & ACC & ASR & ACC & ASR & ACC & ASR & ACC & ASR & ACC & ASR \\ \toprule
        
        \multirow{3}{*}{MNIST} & pre-activate & 96.83 & 19.00 & 96.77 & 27.63 & 97.49 & 19.99 & 83.85 & 15.40 & 97.43 & 24.08 & 97.52 & 28.46 \\

        & Retrain & 96.72 & 96.44 & 96.72 & 99.34 & 97.55 & 96.75 & 82.55 & 97.69 & 97.61 & 96.95 & 97.54 & 98.99 \\

        & FedEraser & 94.86 & 76.56 & 95.01 & 95.65 & 96.78 & 86.97 & 82.53 & 65.38 & 97.26 & 90.36 & 97.28 & 96.90 \\ \midrule
        
        \multirow{3}{*}{CIFAR-10} & pre-activate & 87.29 & 12.91 & 85.16 & 14.83 & 97.52 & 29.80 & 83.78 & 20.02 & 84.84 & 21.55 & 85.23 & 15.85 \\

        & Retrain & 85.00 & 99.84 & 84.97 & 72.45 & 97.53 & 99.20 & 84.64 & 68.48 & 83.90 & 99.32 & 84.71 & 70.30 \\

        & FedEraser & 83.97 & 79.04 & 85.06 & 50.72 & 96.78 & 95.98 & 83.48 & 54.69 & 83.19 & 90.76 & 83.99 & 53.72 \\ \bottomrule
        
    \end{tabular}
   
    \label{tab:vsAggregation}
\end{table*}

Moreover, across all scenarios, the ACC remains within 1\% and at most within 3\% of the model trained in a standard FL setting without malicious clients. This demonstrates that our method preserves the model’s utility on its primary task, thereby reducing the likelihood of detection by the server or other clients based solely on performance metrics during normal federated training process.

\begin{table}[!t]
    \centering
    \caption{Evaluation of \texttt{BadFU} under Dirichlet distribution.}
    \setlength{\tabcolsep}{5pt}
    \resizebox{0.5\textwidth}{!}{
    \begin{tabular}{llcccccc}
        \toprule
        
        \multirow{2}{*}{Dataset} & \multirow{2}{*}{Backdoor} & \multicolumn{2}{c}{Pre-activate} & \multicolumn{2}{c}{Retrain} & \multicolumn{2}{c}{FedEraser} \\ \cmidrule(lr){3-4} \cmidrule(lr){5-6} \cmidrule(lr){7-8}

        & & ACC $\uparrow$ & ASR $\downarrow$ & ACC $\uparrow$ & ASR $\uparrow$ & ACC $\uparrow$ & ASR $\uparrow$ \\ \toprule

        \multirow{2}{*}{MNIST} & BadNet & 97.25 & 27.36 & 97.37 & 98.00 & 96.30 & 90.50 \\

        & Blended & 97.57 & 40.31 & 97.66 & 99.77 & 97.03 & 99.29 \\ \midrule

        \multirow{2}{*}{CIFAR-10} & BadNet & 83.63 & 5.65 & 85.56 & 98.15 & 83.78 & 90.20 \\ 

        & Blended & 84.36 & 8.10 & 83.52 & 92.22 & 83.68 & 83.55 \\
        
        \bottomrule
        
    \end{tabular}
    }
    \label{tab:vsDistribution}
\end{table}

\subsection{Backdoor Injection Algorithm}

To answer \textbf{RQ2}, we conduct an empirical study to evaluate the applicability of \texttt{BadFU} across different backdoor attack techniques. Using identical training and evaluation protocols, we report both the ACC and ASR for each method.
\begin{table}[!t]
    \centering
    \caption{Evaluation of \texttt{BadFU} using Blended under Dominant Class distribution.}
    \setlength{\tabcolsep}{5pt}
    \resizebox{0.5\textwidth}{!}{
    \begin{tabular}{llccccccc}
        \toprule
        \multirow{2}{*}{Dataset} & \multirow{2}{*}{Models} & \multicolumn{2}{c}{Pre-activate} & \multicolumn{2}{c}{Retrain} & \multicolumn{2}{c}{FedEraser} & Benign\\ \cmidrule(lr){3-4}  \cmidrule(lr){5-6} \cmidrule(lr){7-8} \cmidrule(lr){9-9} 
        
        & & ACC $\uparrow$ & ASR $\downarrow$ & ACC $\uparrow$ & ASR $\uparrow$ & ACC $\uparrow$ & ASR $\uparrow$ & ACC $\uparrow$ \\ \toprule
        
        \multirow{2}{*}{MNIST} & SimpleNN & 97.52 & 28.46 & 97.54 & 98.99 & 97.28 & 96.90 & 97.73 \\

        & LeNet-5 & 98.73 & 16.44 & 98.86 & 95.76 & 98.76 & 82.26 & 98.85\\ \midrule

        \multirow{2}{*}{CIFAR-10} & ResNet-18 & 85.23 & 15.85 & 84.71 & 70.30 & 83.99 & 53.72 & 84.81 \\ 

        & VGG-16 & 87.80 & 15.99 & 87.67 & 80.14 & 87.22 & 53.34 & 86.42\\ \bottomrule
        
    \end{tabular}
    }
    \label{tab:FedAvg+Dominant+Blended}
\end{table}

The results are presented in Table~\ref{tab:FedAvg+Dominant+Blended}. Except for SimpleNN on MNIST, all other model–dataset pairs exhibit ASR below 17\% before backdoor activation, indicating that the camouflage samples effectively mask the backdoor behavior during the federated training process. Once the camouflage samples are unlearned, ASR increases sharply across all cases, confirming that the backdoor is successfully activated. Notably, under the Blended backdoor attack, the ASR on CIFAR-10 is slightly lower than that of the BadNet approach, potentially because the Blended method prioritizes trigger stealthiness over effectiveness. Nevertheless, the Blended method still achieves high ASR, posing a significant threat to the security of the federated learning model. The ACC remains comparable to that of the benign setting, suggesting that the model’s primary learning task is largely unaffected and thus unlikely to raise suspicion.

We observe that the difference in model architecture on the same dataset will not have a significant effect on the results. Therefore, for the following subsection, we use SimpleNN on MNIST and ResNet-18 on CIFAR-10 with the BadNet technique for evaluation to answer each RQ.

\subsection{Data Distribution Comparison}

To answer \textbf{RQ3}, we further evaluate the effectiveness of \texttt{BadFU} under an alternative non-IID data distribution commonly used in FL, namely, the Dirichlet distribution. As demonstrated in previous sections, \texttt{BadFU} is effective across various models; hence, for this analysis, we select one representative model per dataset to isolate and evaluate the impact of data distribution on the attack performance.

Combining the results from Table~\ref{tab:FedAvg+Dominant+BadNet} and Table~\ref{tab:vsDistribution}, we observe that both MNIST with SimpleNN and CIFAR-10 with ResNet-18 consistently achieve high ACC across all settings, closely matching the performance of models trained without any malicious clients. This confirms that \texttt{BadFU} preserves the model’s primary learning objectives, regardless of the data partitioning strategy. Meanwhile, the ASR values after unlearning activation are notably higher under the Dirichlet distribution compared to the dominant class setting (all exceeding $83\%$), which may be attributed to the more balanced data spread among clients in the Dirichlet setup, allowing the camouflage samples to integrate more effectively during training. These results demonstrate the effectiveness of \texttt{BadFU} across different non-IID distributions in FL settings. Since \texttt{BadFU} works well on both cases of the Dominant Class distribution and the Dirichlet distribution, we mainly use the dominant class distribution for the following subsections.

\subsection{Aggregation Method Comparison}

To answer \textbf{RQ4}, we evaluate the performance of \texttt{BadFU} under different federated aggregation algorithms. Specifically, we compare the attack effectiveness across three aggregation strategies: FedAvg, FedProx, and FedSGD. In this experiments, dominant class distribution for partitioning client data is used to simulate the typical non-IID setting of FL.

From Table~\ref{tab:vsAggregation}, we observe that pre-activation ASR remains below 30\% across all aggregation strategies, demonstrating that our proposed camouflage strategy effectively conceals the backdoor prior to activation. For the MNIST–SimpleNN pair using the Blended backdoor, FedProx results in a noticeably lower post-activation ASR compared to FedAvg and FedSGD, though it still exceeds 65\%, indicating a strong attack effect. In all other settings, post-activation ASR values are comparable across the three aggregation rules. These results confirm that \texttt{BadFU} remains effective across diverse FL frameworks.

\begin{figure*}[tbp]
    \centering
    \begin{subfigure}[t]{0.3\textwidth}
        \centering
        \includegraphics[width=\linewidth]{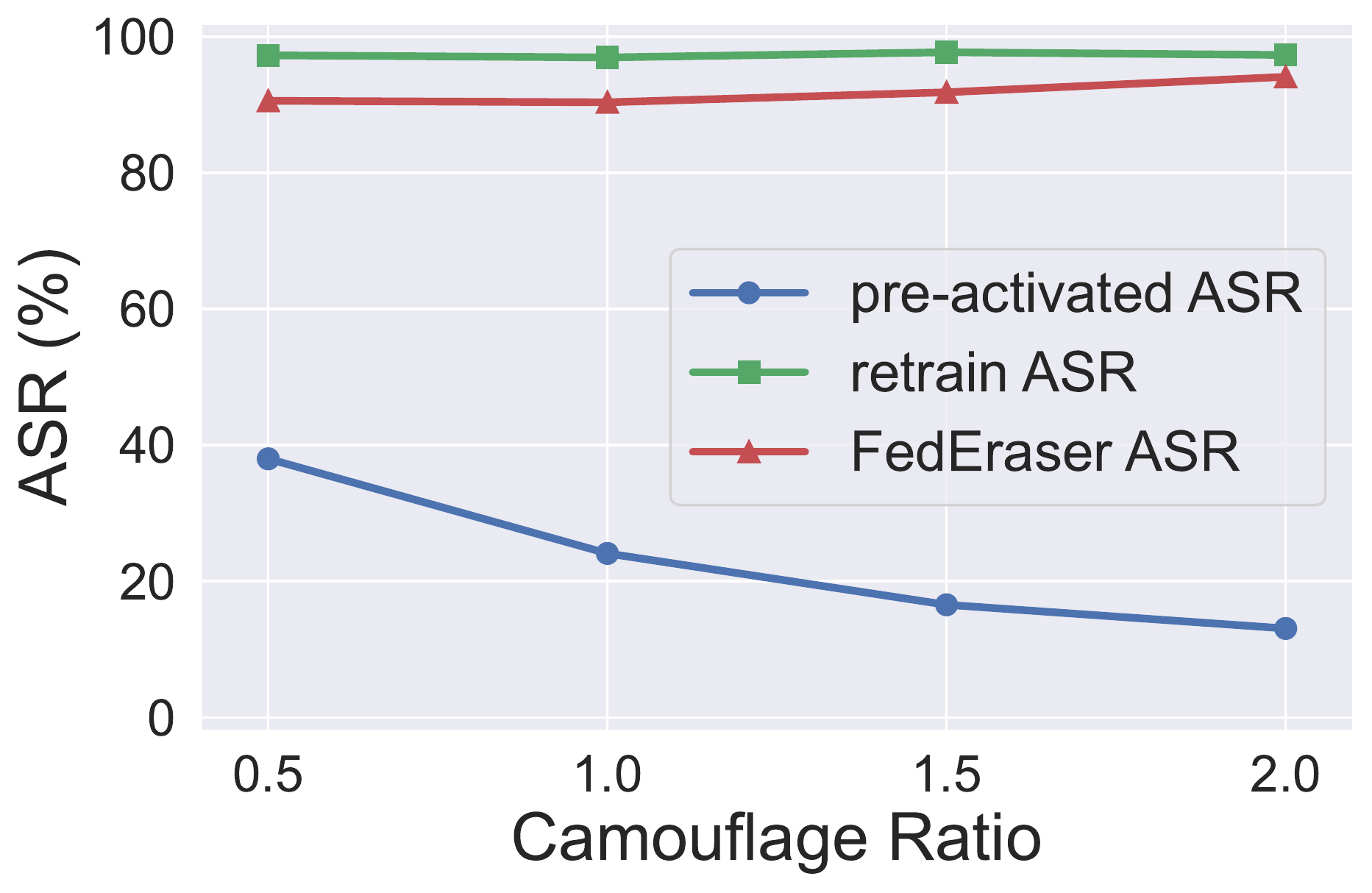}
        \caption{MNIST}
        \label{fig:ratio-mnist}
    \end{subfigure}
    \hspace{1.5cm}
    \begin{subfigure}[t]{0.3\textwidth}
        \centering
        \includegraphics[width=\linewidth]{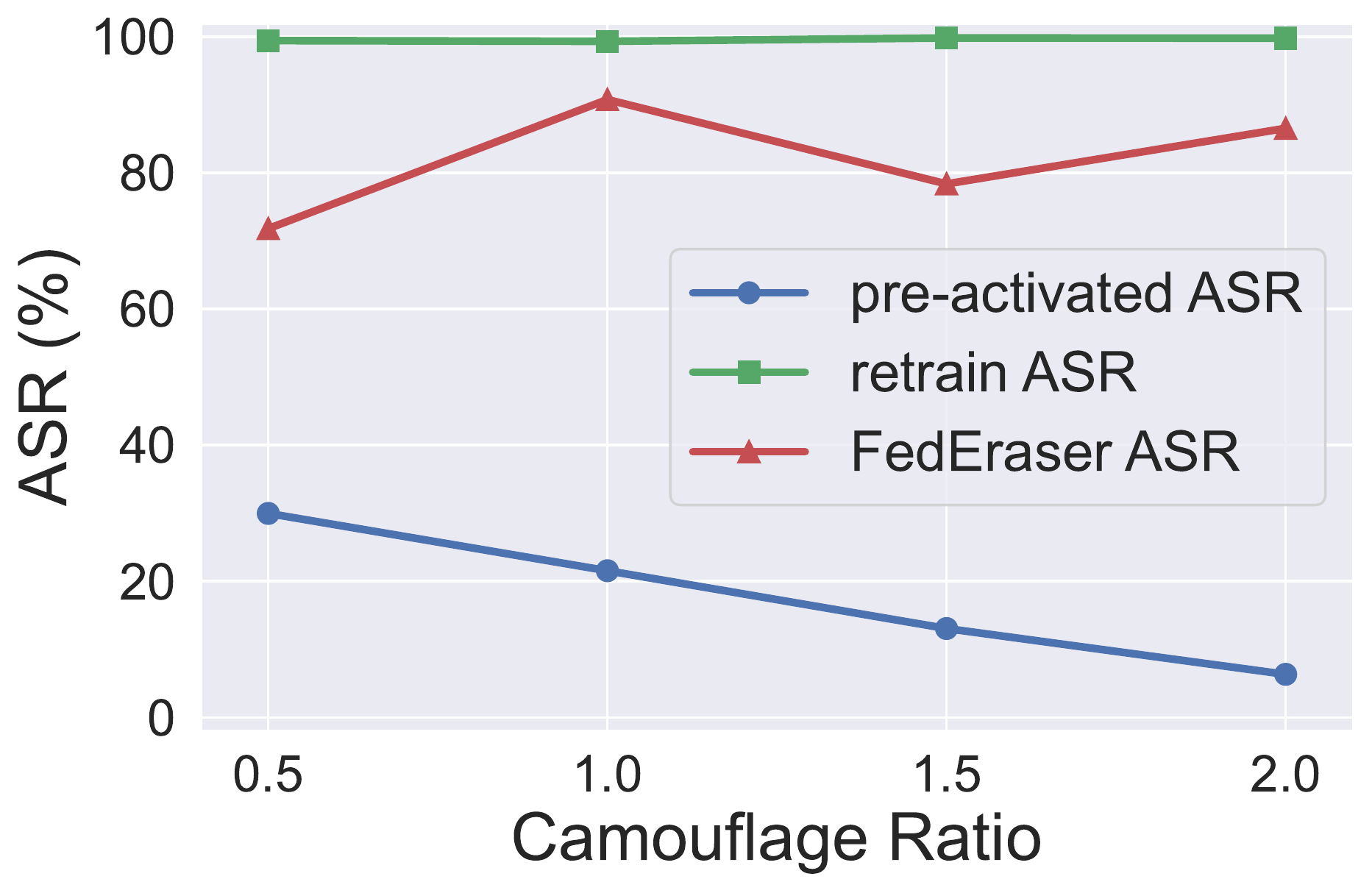}
        \caption{CIFAR-10}
        \label{fig:ratio-cifar}
    \end{subfigure}

    \caption{Attack Success Rate (ASR) versus different camouflage sample/poison sample ratio on MNIST and CIFAR-10.}
    \label{fig:ratio}
\end{figure*}

\subsection{Normal Unlearning Request}

In addition to the above metrics, it is crucial to ensure that normal unlearning requests from benign clients do not unintentionally activate the backdoor. To answer \textbf{RQ5}, we perform a comparative experiment by simulating a benign client submitting a legitimate unlearning request under the FedAvg aggregation algorithm and dominant class distribution.

\begin{table}[!t]
    \centering
    \caption{Performance of \texttt{BadFU} in normal unlearning.}
    \setlength{\tabcolsep}{5pt}
    \begin{tabular}{llcccc}
    
        \toprule
        \multirow{2}{*}{Dataset} & \multirow{2}{*}{Stage} & \multicolumn{2}{c}{BadNet} & \multicolumn{2}{c}{Blended} \\ \cmidrule(lr){3-4} \cmidrule(lr){5-6}
        
        \multicolumn{2}{c}{} & ACC $\uparrow$ & ASR $\downarrow$ & ACC $\uparrow$ & ASR $\downarrow$ \\ \toprule
                
        \multirow{2}{*}{MNIST} & pre-activate & 97.43 & 24.09 & 97.52 & 28.46 \\ 

        & NormalUL & 96.89 & 14.03 & 97.03 & 22.61 \\ \midrule

        \multirow{2}{*}{CIFAR-10} & pre-activate & 84.84 & 21.55 & 85.23 & 15.85 \\

        & NormalUL & 81.29 & 10.22 & 85.25 & 15.24 \\ \bottomrule
        
    \end{tabular}
    \label{tab:BenignUL}
\end{table}

In Table~\ref{tab:BenignUL}, NormalUL refers to the scenario where the global model is unlearned in response to a benign client’s unlearning request. The results show that the ASR before and after this unlearning remains nearly unchanged under the Blended attack setting in \texttt{BadFU}, and even decreases slightly under the BadNet setting. This suggests that unlearning requests from benign clients do not unintentionally activate the backdoor implanted by the malicious client. Therefore, \texttt{BadFU} successfully preserves both the stealthiness of the backdoor and the control of its activation by the adversary.

\subsection{Camouflage/Poison Ratio Sensitivity}

To answer \textbf{RQ6}, we analyze how varying the proportion of camouflage samples relative to backdoor samples affects the effectiveness of \texttt{BadFU}. Specifically, we fix the overall poison ratio and vary the number of camouflage samples to be $0.5\times$, $1\times$, $1.5\times$, and $2\times$ the number of backdoor samples.

Figure~\ref{fig:ratio} illustrates the backdoor ASR under varying camouflage ratios, both before and after federated unlearning. The blue lines represent the ASR before unlearning (i.e., pre-activation), while the green and orange lines correspond to the ASR after unlearning (i.e., post-activation). As shown, the pre-activated ASR decreases steadily from approximately 40\% to below 20\% for both datasets as the camouflage ratio increases from $0.5\times$ to $2\times$. In contrast, the post-activated ASR remains consistently high, above 90\% for MNIST and around 80\% for CIFAR-10—regardless of the camouflage ratio. These results suggest that increasing the number of camouflage samples during local training effectively suppresses visible backdoor behavior in the global model, thereby enhancing stealthiness. Importantly, the backdoor remains potent and is successfully reactivated after unlearning, indicating that the number of camouflage samples does not hinder the eventual attack. This implies that malicious clients can strategically increase camouflage data to strengthen the concealment of the backdoor during federated training but without compromising its effectiveness after federated unlearning.

\subsection{Defense Mechanism}

In this experiment, we address \textbf{RQ7} by evaluating the resilience of \texttt{BadFU} against existing backdoor defenses in federated learning. Specifically, we assess \texttt{BadFU} under two categories of defenses: robust aggregation-based defenses and post-training detection-based defenses. For the former, we consider Median and Trimean~\cite{yin2018byzantine}, which are Byzantine-robust aggregation strategies designed to filter out malicious updates during the training process. For the latter, we apply Neural Cleanse (NC)~\cite{wang2019neural}, a representative detection-based defense that aims to identify potential backdoor triggers in trained models by analyzing class-wise perturbation patterns.

\begin{table}[!t]
    \centering
     \caption{The performance of \texttt{BadFU} under secure aggregation based defenses. Here, BadNet is used as the backdoor technique in \texttt{BadFU}.}
     \setlength{\tabcolsep}{5pt}
    \begin{tabular}{llcccc}
        \toprule
        \multirow{2}{*}{Dataset} & \multirow{2}{*}{Stage} & \multicolumn{2}{c}{Median} & \multicolumn{2}{c}{Trimean} \\ \cmidrule(lr){3-4} \cmidrule(lr){5-6}
        
         & & ACC & ASR & ACC & ASR \\ \toprule
        
        \multirow{2}{*}{MNIST} & pre-activate & 97.51 & 1.14 & 97.62 & 1.42 \\

        & activated & 97.22 & 84.04 & 97.52 & 94.27 \\ \midrule
        
        \multirow{2}{*}{CIFAR-10} & pre-activate & 84.08 & 6.55 & 84.48 & 7.16 \\

        & activated & 82.06 & 96.33 & 83.83 & 76.97 \\ \bottomrule
        
    \end{tabular}
   
    \label{tab:secagg}
\end{table}

\noindent \textbf{Robust Aggregation based Defense.} Table~\ref{tab:secagg} presents the performance of \texttt{BadFU} under robust aggregation defenses during the training stage. It shows that the defense in the training stage to remove the impact of anomalies is unable to defend against \texttt{BadFU} activated (anomalies' impact) in the unlearning stage. After unlearning activation, all ASRs remain high, $96.33\%$ for Median on CIFAR-10 and $94.27\%$ for Trimean on MNIST, demonstrating that the backdoor attack is successfully activated. Notably, the pre-activation ASRs are consistently low (below $10\%$ across all cases), indicating that the backdoor effect in the malicious client's local updates is further masked by the robust aggregation mechanisms. This aligns with the design of \texttt{BadFU}, which does not aim to embed an active backdoor during training, but instead relies on federated unlearning for activation. 

\begin{table}[!t]
\centering
\caption{Evaluation of Neural Cleans (NC) on \texttt{BadFU} on ResNet-18 trained on CIFAR-10. Note that the class with an index value lower than $-2$ is considered as the anomaly class containing a backdoor trigger.}
\setlength{\tabcolsep}{5pt}
\begin{tabular}{lccc}
    \toprule
    \multirow{2}{*}{Label} & \multicolumn{3}{c}{Model} \\ \cmidrule(lr){2-4} 
    & Clean & Backdoor & \texttt{BadFU} \\ \toprule
    \textbf{0} & 0.53 & -0.71 & -0.12 \\ 
    \textbf{1} & -1.82 & 1.51 & -1.33 \\ 
    \textbf{2} & -0.38 & -0.37 & 0.64 \\ 
    \textbf{3} & 0.82 & 0.48 & 0.40 \\ 
    \textbf{4} & -0.51 & -0.64 & -0.71 \\ 
    \textbf{5} & 3.72 & 1.65 & 2.13 \\ 
    \textbf{6} & -2.24 & -1.39 & -0.85 \\ 
    \textbf{7} & 0.38 & 0.37 & 0.12 \\ 
    \textbf{8} & 0.46 & 0.43 & -0.86 \\ 
    \textbf{9} & -1.82 & -1.00 & 0.51 \\
    \bottomrule
\end{tabular}
\label{tab:nc}
\end{table}

\noindent \textbf{Detection-based Defense.} NC considers a class to contain a backdoor trigger if its index is lower than $-2$. Table~\ref{tab:nc} presents the NC indices of all classes in the global model, ResNet-18, after federated training on CIFAR-10, with class $0$ set as the target in this experiment. We observe that NC does not perform well in detecting backdoor in models trained under the federated learning paradigm. Specifically, in the plain backdoor scenario, where a malicious client directly injects a backdoor during federated training, NC fails to confidently detect the presence of the backdoor. Even in the clean model, NC may misclassify a benign class (e.g., label $6$ with an index of $-2.24$) as being backdoored. Furthermore, we find that the NC index exhibits instability across repeated experiments. A possible explanation is that NC was originally designed for centralized training, where the backdoor pattern is typically more prominent. In contrast, federated models are generally less overfit to the training data, making backdoor patterns more subtle~\cite{sun2024understanding}.

These experimental results highlight a critical gap in existing studies of FL and federated unlearning: current backdoor defenses in FL are insufficient to prevent post-unlearning backdoor activation, e.g., \texttt{BadFU}, underscoring the urgent need for robust and secure unlearning mechanisms to ensure the integrity of unlearned models.

\subsection{Unlearning Method}

In this experiment, we evaluate the applicability of \texttt{BadFU} on two additional approximate federated unlearning methods, specifically, the influence function-based unlearning of FedU and the certified unlearning of SIFU, to address \textbf{RQ8}. For FedU, we use the MNIST dataset with a SimpleNN architecture, and for SIFU, we conduct experiments on CIFAR-10 using ResNet-18. In the SIFU setting, we constrain the unlearning epoch to half of the total global communication rounds, following the configuration in the original paper~\cite{fraboni2024sifu}. For FedU, we adopt the parameter settings from~\cite{wang2024fedu}, with the influence function weight set to $\lambda = 0.5$ and the utility preservation coefficient set to $\beta = 0.5$.

\begin{table}[t]
    \centering
    \caption{Evaluation of \texttt{BadFU} across different federated unlearning methods.}
    \setlength{\tabcolsep}{5pt}
    \begin{tabular}{lllcc}
        \toprule
        \multicolumn{3}{c}{Unlearning Methods} & ACC & ASR \\ \toprule
        
        \multirow{2}{*}{MNIST} & \multirow{2}{*}{FedU} & pre-activate & 97.60 & 24.11 \\ 

        & & activated & 97.65 & 60.12 \\ \hline

        \multirow{2}{*}{CIFAR-10} & \multirow{2}{*}{SIFU} & pre-activate & 83.70 & 14.69 \\ 

        & & activated & 83.85 & 85.59 \\ \toprule
        
    \end{tabular}    
    \label{tab:unlearning}
\end{table}

\begin{table}[t]
    \centering
    \caption{Evaluation of \texttt{BadFU} on CIFAR-100, where $\star$ and $\star\star$ represents the ratio between the camouflage samples and poison samples is 1:1 and 2:1.}
    \setlength{\tabcolsep}{5pt}
    \begin{tabular}{cccccc}
        \toprule
        \multicolumn{2}{c}{\multirow{2}{*}{Backdoor Attacks}} & \multicolumn{2}{c}{$\star$} & \multicolumn{2}{c}{$\star\star$} \\ \cmidrule(lr){3-4} \cmidrule(lr){5-6}
        
        \multicolumn{2}{c}{} & ACC & ASR & ACC & ASR \\ \toprule
        
        \multirow{3}{*}{BadNet} & pre-activate & 61.02 & 39.62 & 61.81 & 26.84 \\ 

        & Retrain & 59.99 & 99.62 & 59.57 & 99.21 \\ 

        & FedEraser & 59.60 & 72.33 & 60.80 & 66.10 \\ \hline

        \multirow{3}{*}{Blended} & pre-activate & 61.21 & 39.43 & 61.39 & 23.67 \\ 

        & Retrain & 61.64 & 86.30 & 63.30 & 88.80 \\ 

        & FedEraser & 59.58 & 53.52 & 59.48 & 54.81 \\ \bottomrule
        
    \end{tabular}
    
    \label{tab:cifar100}
\end{table}

Table~\ref{tab:unlearning} demonstrates that \texttt{BadFU} can be effectively applied to both FedU and SIFU, achieving relatively high backdoor attack success rates (ASR). In the case of SIFU, the pre-activation ASR remains low at $14.69\%$, but increases significantly to $85.59\%$ after the unlearning process is triggered. For FedU, the ASR exceeds $60\%$ following unlearning activation, which is sufficient to pose a serious threat to the integrity of the global model. Furthermore, the attacker can achieve even higher ASR in the FedU setting by injecting more poisoned samples during the federated training stage. These poisoned samples can remain undetected during training, as their effects are masked by camouflage samples. In summary, \texttt{BadFU} can be generalized to other federated unlearning methods, reinforcing the practicality and severity of security threats inherent in existing federated unlearning mechanisms.

\subsection{More Complex Dataset}

To further evaluate the scalability of our approach on more complex learning tasks, we conduct additional experiments on CIFAR-100, which comprises 100 fine-grained object categories. As shown in Table~\ref{tab:cifar100}, \texttt{BadFU} continues to perform effectively in this more challenging setting. However, the camouflage effect is notably weaker on CIFAR-100. For both backdoor attack strategies, the pre-activation ASR exceeds 39\% when the number of camouflage samples equals the number of backdoor samples, which is significantly higher than in previous experiments. To enhance the camouflage effect, the attacker can increase the ratio of camouflage to backdoor samples. Under the configuration where the number of camouflage samples is twice that of the backdoor samples, the pre-activation ASRs drop below 30\%, while the benign accuracy and the ASRs after unlearning activation remain comparable to those in the 1:1 setting. These results suggest that although the camouflage effect of \texttt{BadFU} is somewhat diminished in complex learning tasks, it still remains a viable and effective strategy for attacking federated learning systems.

\section{Discussion and Limitations}

\noindent\textbf{Cross-silo Setting.} Although our experiments are conducted in a cross-silo FL environment with non-IID data distribution, the proposed method is not inherently limited to this setting. In principle, it can be extended to other FL scenarios, such as cross-device settings characterized by more clients and greater data heterogeneity. These environments, however, pose additional challenges for the attacker, including lower client participation rates and stronger counter-effects from benign clients. To successfully execute a backdoor attack via federated unlearning under such conditions, the proposed method would need to be augmented with more advanced data poisoning strategies—for example, the attack-of-the-tails technique~\cite{wang2020attack}.

\noindent\textbf{FL Frameworks.} \texttt{BadFU} is, in principle, also applicable to FL settings employing alternative aggregation mechanisms such as Trimmed-Mean, FedMedian, or other Byzantine-robust approaches~\cite{yin2018byzantine}. This is because our attack targets the backdoor activation after model training, rather than injecting the backdoor during the training process. We acknowledge that these robust aggregation methods are designed to diminish the influence of malicious client updates, thereby increasing the difficulty of implanting a backdoor into the global model. Nevertheless, our approach can be adapted to such settings by incorporating more advanced data poisoning techniques. It is worth emphasizing that this work represents the first attempt to exploit federated unlearning as a vector for backdoor attacks. Thus, our primary objective is to expose the security vulnerabilities of existing federated unlearning mechanisms, rather than to comprehensively attack all FL frameworks.

\noindent\textbf{Backdoor Defense Mechanisms.} Existing defenses against backdoor attacks can be broadly categorized into two types: defenses during training and post-training detection. Training-time defenses, such as Byzantine-robust aggregation methods (including Median~\cite{yin2018byzantine}, Trimean~\cite{yin2018byzantine}, Krum~\cite{blanchard2017machine}, and FLTrust~\cite{cao2021fltrust}), aim to filter out or down-weight suspicious local updates contributed by potentially malicious clients. In contrast, our approach introduces a novel threat in which the backdoor remains dormant and largely indistinguishable from benign behavior during training, only becoming active during the unlearning process. This characteristic allows the backdoor to evade training-time defenses effectively.
For post-training detection mechanisms, such as Neural Cleanse (NC)~\cite{wang2019neural}, our approach conceals the impact of the backdoor trigger, thereby reducing detection efficacy. Moreover, in our experiments, we find that detection-based methods often exhibit degraded performance in FL settings, particularly under the non-IID data assumption, which further limits their practical effectiveness.

\noindent\textbf{Mitigation against \texttt{BadFU}.} Since \texttt{BadFU} launches the backdoor attack via federated unlearning after the model has been trained, traditional defenses targeting malicious updates during training are not directly effective. Nonetheless, the underlying principle of filtering out malicious behavior can potentially be adapted to defend against \texttt{BadFU}. Specifically, during the federated unlearning process, the server could analyze whether unlearning requests exhibit suspicious patterns by comparing the characteristics of the data marked for unlearning against the distribution of benign training data. If a request is determined to be malicious, countermeasures such as request rejection or data purification can be applied.

Another promising direction is to holistically monitor the global model’s behavior, both during training and after unlearning, using advanced backdoor detection techniques. If a backdoor is detected, the model can be purified prior to redistribution to local clients, thereby limiting its impact.

Lastly, differential privacy (DP) has been investigated to enhance both data privacy and model robustness in FL. Current DP based defense mechanisms, e.g., DP-FL~\cite{huang2020dp}, focus on the training stage and hence cannot defend \texttt{BadFU} effectively. However, in principle, the server could apply DP mechanisms during unlearning by injecting differentially private noise, thereby limiting the influence of the unlearned samples on the updated model. However, the introduction of noise inevitably degrades the utility of the global model. As such, achieving an optimal trade-off between model utility and robustness against backdoor attacks remains a significant and ongoing challenge.

\noindent\textbf{Limitations.} This work has two main limitations: \textit{i}) its applicability to generative tasks and other data modalities, and \textit{ii}) its effectiveness on large and complex datasets. Our study primarily focuses on classification tasks, without investigating how \texttt{BadFU} performs in generative settings such as image or text generation. Additionally, the experiments are conducted exclusively on image datasets, and do not explore other data types such as text. Another limitation lies in the scalability of \texttt{BadFU} to more complex datasets. As demonstrated in our experiments, the camouflage effect of \texttt{BadFU} becomes weaker on more complex datasets such as CIFAR-100.

We believe that extending \texttt{BadFU} to other learning tasks, data modalities, and larger datasets presents promising directions for future work. Such efforts would not only deepen the understanding of the vulnerabilities inherent in federated unlearning mechanisms, but also contribute to the development of more robust and secure federated learning systems. 

\section{Conclusion}

This paper introduces \texttt{BadFU}, a novel attack framework that leverages federated unlearning to stealthily implant and subsequently activate a backdoor in the global model within cross-silo federated learning. By embedding a mixture of backdoor and camouflage samples into the local training data of a malicious client, \texttt{BadFU} preserves a low attack success rate and high benign accuracy during the federated training phase, prior to backdoor activation, while achieving high attack success rates after an unlearning request is issued. Comprehensive experiments across benchmark datasets, model architectures, FL frameworks, and non-IID environments validate the feasibility and effectiveness of the proposed approach. Furthermore, we demonstrate that existing backdoor defenses in federated learning are insufficient to counter \texttt{BadFU}. These findings expose critical security vulnerabilities in current federated unlearning protocols, which may unintentionally enable potent backdoor attacks.

\bibliographystyle{IEEEtran}
\bibliography{reference}

\section*{Appendix}
\section{Related Work}
We provide a detailed introduction of related works of federated learning and machine unlearning as follows.
\subsection{Federated Learning}
Federated learning is a collaborative learning paradigm that enables multiple clients to jointly train a machine learning (ML) model without sharing their raw data~\cite{yin2021comprehensive, zhang2021survey}. Originally introduced by Google~\cite{konevcny2016federated}, FL addresses privacy concerns by keeping data local to each client. One of the earliest approaches, FedSGD~\cite{mcmahan2017communication}, asks each client to perform a single epoch local gradient computation before uploading it to a central server for aggregation. This approach is straightforward, but it can incur high communication costs. Thus, FedAvg~\cite{mcmahan2017communication} was proposed to reduce communication overhead by allowing clients to perform multiple epoch local updates on its dataset and only communicate the resulting model parameters. In addition to FedSGD and FedAVG, there are other FL frameworks, such as FedProx~\cite{li2020federated}, Median~\cite{yin2018byzantine}, and Trimmed-Mean~\cite{yin2018byzantine} that are proposed to mitigate different issues in federated training process.

In FL, one of the main challenges is handling heterogeneous data distribution across clients~\cite{zhao2018federated}. In real-world deployments, client datasets often do not follow the same distribution as the global dataset, leading to non-independent and identically distributed (non-IID) data. This heterogeneity can severely degrade model performance and training stability. Techniques such as FedProx~\cite{li2020federated} have been specifically designed to mitigate the adverse effects of such data imbalance.

As FL is a collaborative paradigm, it is almost inevitable that some clients will eventually withdraw and require unlearning, demanding that their data contributions be removed from the collaboratively trained model~\cite{halimi2022federated}. This introduces an entirely new requirement for federated unlearning.

\subsection{Machine Unlearning}

Machine unlearning has gained attention as a method for removing the influence of specific data from trained models, driven in part by privacy regulations such as the GDPR~\cite{mantelero2013eu} and the CCPA~\cite{goldman2020introduction}. Existing approaches of machine unlearning can mainly be categorized into exact unlearning and approximate unlearning methods~\cite{zhang2023review}. Exact unlearning approaches aim to remove data influence by efficiently retraining the model. A representative approach is SISA~\cite{bourtoule2021machine}, which partitions the dataset into different shards, trains models on each shard, and aggregates the predictions of these models together during the inference stage. When an unlearning request is raised, it first locates the shards where the sample needs to be deleted, retrains only these shards, and then aggregates all the models again to achieve fast retraining. 
Unlike exact unlearning methods, approximate unlearning focuses on removing the influence of a specific training data by directly modifying the parameters of the trained model. Under the approximate unlearning framework, approaches utilize gradients or influence functions to estimate the influence of selected samples and then subtract it from the model~\cite{warnecke2021machine}. Because of directly removing the information of the sample from the model, approximate unlearning usually achieves far more efficiency than exact unlearning~\cite{liu2025rethinking,ye2023reinforcement}.

In the FL setting, machine unlearning becomes more challenging due to iterative and incremental model updates~\cite{wu2022federated}. The influence of a specific data sample can spread across all clients' local models through aggregation. FedEraser~\cite{liu2021federaser} is one of the most well-known approaches for federated unlearning. It operates orthogonally to the FL system, requiring only that each client periodically records local model updates. When an unlearning request is received, each client retrains the local model and sends the recorded and updated parameters to the server for aggregation every few rounds. KNOT~\cite{su2023asynchronous} is another retraining-based federated unlearning method using asynchronous cluster aggregation. FedU~\cite{wang2024fedu} is an influence function based federated unlearning method. In FedU, the client calculates the influence of deleted samples, subtracts it from the local model, and then performs the utility-preserving training on a mini batch sampled from the remaining local dataset; all other clients do the local training on a mini batch sampled from their own local dataset. SIFU~\cite{fraboni2024sifu} saves the sensitivity of the clients during the training. When receiving an unlearning request, it first adds noise to the model to achieve the unlearning effect; then does a few global communication rounds of normal training to preserve the utility. 
\end{document}